\documentclass[12pt]{article}
\usepackage{amsfonts}
\usepackage[utf8]{inputenc}
\usepackage{bm}
\usepackage{amsmath}
\usepackage{amssymb}
\usepackage{epsfig}
\usepackage{enumitem}
\usepackage{cite}
\usepackage[bookmarks=true,colorlinks=true,linkcolor=black,citecolor=orange,urlcolor=orange,bookmarksnumbered]{hyperref}
\usepackage{cancel}

\usepackage[dvipsnames]{xcolor}

\setlist[enumerate]{%
wide =0.5\parindent,
listparindent=0pt%
}

\newcommand{\bmat}{\left(\begin{array}}
\newcommand{\emat}{\end{array}\right)}

\def\gtrsim{\mathrel{\raise.3ex\hbox{$>$\kern-.75em\lower1ex\hbox{$\sim$}}}}

\def\a{\alpha}

\def\ov{\overline}
\def\un{\underline}

\def\-{\hphantom{-}}
\def\ov{\overline}
\def\s2{\frac{1}{\sqrt2}}

\def\mg{m_{3/2}}
\def\mg2{m^2_{3/2}}

\def\Dsl{\,\raise.15ex\hbox{/}\mkern-13.5mu D} %this one can be subscripted

\def\be{\begin{equation}}
\def\ee{\end{equation}}
\def\bea{\begin{eqnarray}}
\def\eea{\end{eqnarray}}

\newcommand{\nn}{\nonumber}

\begin{document}

%\tableofcontents
%\newpage

\pagestyle{plain}

%----------------------------------------------------------------------%
%  numbering equations with section number
%----------------------------------------------------------------------%
\makeatletter
\@addtoreset{equation}{section}
\makeatother
\renewcommand{\theequation}{\thesection.\arabic{equation}}
%----------------------------------------------------------------------%
%  title page
%----------------------------------------------------------------------%
\pagestyle{empty}
\begin{center}
\ \

\vskip .5cm

\LARGE{\bf Non-Relativistic Limits of Bosonic and Heterotic Double Field Theory} \\[10mm]

\vskip 0.3cm

\large{Eric Lescano and David Osten
 \\[6mm]}

{\small Institute for Theoretical Physics (IFT), University of Wroc\l aw, \\
pl. Maxa Borna 9, 50-204 Wroc\l aw,
Poland\\ [.01 cm]}

{\small \verb"{eric.lescano,david.osten}@uwr.edu.pl"}\\[1cm]

\small{\bf Abstract} \\[0.5cm]

 \end{center}

The known stringy non-relativistic (NR) limit of the universal NS-NS sector of supergravity has a finite Lagrangian due to non-trivial cancellations of divergent parts coming from the metric and the $B$-field. We demonstrate that in Double Field Theory (DFT) and generalised geometry these cancellations already happen at the level of the generalised metric, which is convergent in the limit $c \rightarrow \infty$, implying that the NR limit can be imposed before solving the strong constraint. We present the $c$-expansion of the generalised metric, which reproduces the Non-Riemannian formulation of DFT at the (finite) leading order, and the $c$-expansion of the generalised frame, which contains divergences. We also extend this approach to the non-Abelian gauge field of Heterotic DFT assuming a convergent expansion for the O$(D,D+n)$ generalised metric. From this proposal, we derive a novel $c$-expansion for the bosonic part of the heterotic supergravity which is, by construction, compatible with O$(D,D)$-symmetry.

%\newpage
%\newpage
%----------------------------------------------------------------------%
%  Resetting of counters 

%----------------------------------------------------------------------%
\setcounter{page}{1}
\pagestyle{plain}
\renewcommand{\thefootnote}{\arabic{footnote}}
\setcounter{footnote}{0}
%----------------------------------------------------------------------%
%  Paper begins
%----------------------------------------------------------------------%
\newpage
\tableofcontents 
\section{Introduction}

Recently, a lot of progress has been made in understanding the supergravity limit of non-relativistic (NR) string theory (ST) \cite{Gomis:2000bd,Danielsson:2000gi,Bergshoeff:2018vfn,Harmark:2019upf,Bergshoeff:2019pij,Yan:2019xsf,Julia:1994bs,Bergshoeff:2018yvt,Kluson:2018egd} considering the universal NS-NS sector of ST \cite{Bergshoeff:2015uaa,Bergshoeff:2021bmc,Bidussi:2021ujm}. For heterotic supergravity, the construction of the bosonic part of the theory was recently developed in \cite{BergshoeffRomano}, while the fermionic sector was studied previously in \cite{Bergshoeff:2021tfn}. All these setups are in agreement with the worldsheet formulation of the NRST, where typically the worldsheet action requires an interesting interplay between the vielbein and the NS-NS two-form field in order to lead to a cancellation of the divergences when the NR limit of the relativistic string action is taken. The construction of the effective supergravity prescription follows the same logic, and it starts by considering a $c$-expansion for the ten-dimensional vielbein $\hat e_{\mu}{}^{\hat a}$, i.e.,
\bea
\hat e_{\mu}{}^{a} = c \ \tau_{\mu}{}^{a} \, , \qquad \hat e_{\mu}{}^{a'} =  e_{\mu}{}^{a'} \, ,
\eea
where $\mu,\nu=1,\dots,10$, $a=0,1$ and $a'=2,\dots,9$. When one tries to use this expansion on the NS-NS Lagrangian, taking the limit $c\rightarrow \infty$ leads to divergent gravitational theory. However, a particular expansion for the $B$-field (which is also compatible with the worldsheet formulation of NRST) can be used to construct a well behaved supergravity,
\bea
\hat B_{\mu \nu} & = & - c^2 \epsilon_{a b} \tau_{\mu}{}^{a} \tau_{\nu}{}^{b} + b_{\mu \nu} \, .
\label{expansionBintro}
\eea
The cancellation, at the supergravity level, is straightforward but not automatic, and requires to expand the full supergravity Lagrangian to prove that the divergent contribution from the Ricci scalar can be regulated by the $H^2$ term. While the dilaton has the $c$-expansion 
\bea
\hat{\Phi} = \ln(c) + \phi \, ,
\eea
s.t. the divergences in the measure $\sqrt{- \hat{g}} e^{- 2 \hat{\Phi}}$ of the supergravity action cancel.

\paragraph{The NR limit of bosonic DFT.} The effect of T-duality as a fundamental symmetry of NRST has been studied in several works. This symmetry maps a space-like longitudinal direction to a null one, underlying a discrete lightcone quantisation in the NRST. In this work, we rewrite the NR supergravity in the language of Double Field Theory (DFT) \cite{Siegel1,Siegel2,DFT1,DFT2,DFT3,DFT4} and generalised geometry, which are (double) geometries in which the low energy limit of ST can be written in a T-duality invariant way (see \cite{ReviewDFT1,ReviewDFT2,ReviewDFT3} and the second lecture of \cite{ReviewDFT4} for reviews on this topic). For the bosonic case, the generalised metric $\hat{\cal H}(\hat{g},\hat{B})$ contains divergences in its components, but when one considers the $B$-field expansion (\ref{expansionBintro}), the components of $\hat{\cal H}(\hat g,\hat B)$ are all convergent, i.e.,
\bea
\hat{\cal H}(\hat g,\hat B)= {\cal H}^{(0)}(\tau,e,b) + \mathcal{O}\left(\frac{1}{c^2}\right) \, , 
\eea
and the full NR bosonic supergravity can be extracted directly from the DFT Lagrangian, considering only ${\cal H}^{(0)}(\tau,e,b)$ given by
\be  
{{\cal H}}^{(0)}_{M N} (\tau,e,b) =  
\left(\begin{matrix} h^{\mu \nu} &  \epsilon_{a b} \tau_{\nu}{}^{b} \tau^{\mu a}  - b_{\rho \nu} h^{\rho \mu} \\ 
\epsilon_{a b} \tau_{\mu}{}^{b} \tau^{\nu a}  - b_{\rho \mu} h^{\rho \nu} &  h_{\mu \nu} + b_{\rho \mu} h^{\rho \sigma} b_{\sigma \nu} - 2 \epsilon_{c d}  \tau_{(\mu|}{}^{d} b_{\sigma |\nu)} \tau^{\sigma c}
\end{matrix}\right) 
\ee
which is a so-called \textit{non-Riemannian generalised metric} \cite{PM}. For the flat space, the Gomis-Ooguri background \cite{Gomis:2000bd}, this had been demonstrated in \cite{Ko:2015rha} and for arbitrary backgrounds in \cite{NRDFT2}. The proporties of this Non-Riemmanian limit of DFT have been studied in \cite{NRDFT0,NRDFT1,NRDFT2,NRDFT3,Ko:2015rha}. Motivated by this result, a similar observation was also made in the case of exceptional field theory for Newton-Cartan limits of M-theory \cite{Blair:2021waq}.

For the bosonic case, we extend the existing discussion by presenting the $c$-expansion of DFT both in the generalised metric and generalised frame formalism:
% \begin{align}
%     \begin{array}{ccc}
%       \textrm{supergravity }(\hat{g},\hat{B},\hat{\Phi})   &  \longleftrightarrow & \textrm{bosonic DFT }(\hat{\cal H},\hat{d})\\
%        \downarrow  & & \downarrow \\
%        \textrm{NR limit }(\tau,e,b,\phi) & \longleftrightarrow & \textrm{O}(D,D)\textrm{-invariant NR limit }({\cal H}^{(0)},d)
%     \end{array} \nonumber
% \end{align}
\begin{align}
    \begin{array}{ccccc}
      \textrm{supergravity }   &  \longleftrightarrow & \textrm{bosonic DFT }& &\\
      (\hat{g},\hat{B},\hat{\Phi})& & (\hat{\cal H},\hat{d}) \text{ or } (\hat{ E},\hat{d}) & & \\
       \downarrow  & & \downarrow & \searrow & \\
       \text{\textit{non}-finite $c$-expansion} &  \longleftrightarrow & \text{finite $c$-expansion} & & \text{\textit{non}-finite $c$-expansion} \\
        (\hat{g},\hat{B},\hat{\Phi}) & & (\hat{\mathcal{H}},\hat{d}) & & (\hat{E},\hat{d}) \\
       \downarrow  & & \downarrow & & \downarrow\\
       \textrm{finite NR-supergravity action} & \longleftrightarrow & \text{Non-Riemannian DFT}  & & \text{finite NR-DFT action} \\
       (\tau,e,b,\phi) & & ({\cal H}^{(0)},d) & &
    \end{array} \nonumber
\end{align}
In this diagram, the upper downwards arrow denotes the $c$-expansion of the fundamental variables of the theory, wheres the lower downwards-arrow denotes the limit $c\rightarrow \infty$. The $\leftrightarrow$-arrow means
\bea
& \rightarrow &\textrm{: doubling the coordinates and construct T-duality invariant multiplets,} \nn \\
&\leftarrow &\textrm{: breaking the duality group.} \nn
\eea
The endpoints of this diagram (the first and last row) have been presented before. We argue that the $c$-expansion (the middle row) is also relevant for two reasons:
\begin{itemize}
    \item It shows that the $c$-expansion for the generalised frame variables $\hat{E}$ is not finite. This becomes relevant when considering the NR-limit of DFT with $\alpha^\prime$-corrections where there is no generalised metric formulation \cite{Hohm:2016yvc} in general. See section \ref{chap:HigherDer} for more details.
    
    \item $\frac{1}{c^2}$-corrections are relevant for the study of the string world-sheet in DFT and generalised geometry. For example, in analogy to the point particle one also has to consider $\frac{1}{c^2}$-corrections to $\tau$, the string Bargmann field \cite{Bergshoeff:2019pij,Hartong:2021ekg,Hartong:2022dsx}.
\end{itemize}

\paragraph{The NR limit of heterotic DFT.} We start embedding the proposal given by Bergshoeff and Romano \cite{BergshoeffRomano} in DFT. In this case, the generalised metric has divergent contributions
\bea
\hat{\cal H} = c^4 {\cal H}^{(4)} + c^2 {\cal H}^{(2)} + {\cal H}^{(0)} + \mathcal{O}\left(\frac{1}{c^2}\right) \, . 
\eea
Nevertheless, the limit $c\rightarrow \infty$ is well defined at the supergravity level. While in bosonic DFT the  $c\rightarrow \infty$ limit can be taken before solving the strong constraint and breaking the duality group, this is not possible in this $c$-expansion of heterotic supergravity.
\begin{align}
    \begin{array}{ccc}
      \textrm{heterotic supergravity }(\hat{g},\hat{B},\hat{A},\hat{\Phi})   &  \longleftrightarrow & \textrm{heterotic DFT }(\hat{\cal H},\hat{d})\\
       \downarrow  & & \cancel{\downarrow} \\
       \textrm{NR limit }(\tau,e,b,a,\alpha,\phi) & \cancel{\longleftrightarrow} & \textrm{O}(D,D)\textrm{-invariant NR limit }({\cal H}^{(0)}, {\cal H}^{(2)}, {\cal H}^{(4)},d)
    \end{array} \nonumber
\end{align}
For this reason, we propose a \textit{new expansion for the heterotic fields},
\begin{align}
    \hat{g}_{\mu \nu} &= c^2 {\tau_\mu}^a {\tau_\nu}^b \eta_{ab} + h_{\mu \nu} \nonumber\\
    \hat{B}_{\mu \nu} &= c^2 {\tau_\mu}^a {\tau_\nu}^b \beta_{ab} + b_{\mu \nu} \label{eq:HeteroticExpansionIntro}\\
    \hat{A}_{\mu} &= c \ {\tau_\mu}^a \alpha_a^i + \frac{1}{c} a_\mu^i \nn
\end{align}
which after imposing
\begin{align}
    \alpha^i_a &\equiv \alpha_+^i, \quad \beta_{ab} = \epsilon_{ab}, \qquad \text{or} \qquad \alpha^i_a \equiv \alpha_-^i, \quad \beta_{ab} = -\epsilon_{ab} \label{eq:NewExpansionSolutionIntro}
\end{align}
leads to a finite generalised metric in heterotic DFT
\begin{equation}
 \hat{\cal H}(\hat g, \hat b, \hat A) ={\mathcal{H}}^{(0)}(\tau,h,b,\alpha,a,\phi) + \mathcal{O}\left(\frac{1}{c}\right) \, .
\end{equation}
Therefore, in this case we find
\begin{align}
    \begin{array}{ccc}
      \textrm{heterotic supergravity }(\hat{g},\hat{B},\hat{A},\hat{\Phi})   &  \longleftrightarrow & \textrm{heterotic DFT }(\hat{\cal H},\hat{d})\\
       \downarrow  & & \downarrow \\
       \textrm{NR limit }(\tau,e,b,a,\alpha,\phi) & \longleftrightarrow & \textrm{O}(D,D)\textrm{-invariant NR limit }({\cal H}^{(0)},d) \, .
    \end{array} \nonumber
\end{align}

\paragraph{Organisation of the paper.} In section \ref{Review} we give a quick review to the supergravity formulation of NRST. We set up our notation and we present the fields, the symmetry transformations and the action. Then in section \ref{Bosonic} we reformulate the bosonic supergravity in the standard language of DFT. We start by constructing the generalised frame, and from it we construct the generalised metric. After considering the $B$-field expansion, we arrive to convergent expansion for the generalised metric. Then we compare these results with the non-Riemannian formulation of DFT, finding agreement. At the end of this section we discuss the implications of the divergent terms in the generalized frame formulation for the higher-derivative structure of DFT. Then, in section \ref{Heterotic} we follow a similar procedure but including the heterotic gauge field in a duality covariant way. Following the prescription of \cite{BergshoeffRomano} leads to a finite supergravity but the components of the generalised metric are divergent. Therefore, in the final part of this section we give an alternative proposal for the expansion of heterotic supergravity, which leads to a finite generalised metric. Finally, we conclude with a discussion in section \ref{Discussion}. 

\section{Review: The supergravity expansion}
\label{Review}

We start by splitting the flat index $\hat a=(a,a')$ where $a,b, ...=0,1$ are the transverse directions and $a^\prime,b^\prime,...=2,\dots,9$. The supergravity vielbein, $\hat e_{\mu}{}^{\hat a}$, splits into
\bea
\hat e_{\mu}{}^{a} = c \ \tau_{\mu}{}^{a} \, , \quad \hat e_{\mu}{}^{a'} =  e_{\mu}{}^{a'} \, .
\eea
The inverses of these fields are
\bea
\hat e^{\mu}{}_{a} = \frac{1}{c} \tau^{\mu}{}_{a} \, , \quad \hat e^{\mu}{}_{a'} =  e^{\mu}{}_{a'} \, ,
\eea
and all these quantities obey the following Newton-Cartan relations,
\begin{align}
  \tau_{\mu}{}^{a} e^{\mu}{}_{a'} & = \tau^{\mu}{}_{a} e_{\mu}{}^{a'} = 0 \, , \qquad  e_{\mu}{}^{a'} e^{\mu}{}_{b'} = \delta^{a'}_{b'} \, , \\
  \tau_{\mu}{}^{a} \tau^{\mu b} & = \eta^{a b} \, , \qquad \tau_{\mu}{}^{a} \tau^{\nu}{}_{a} + e_{\mu}{}^{a'} e^{\nu}{}_{a'} = \delta_{\mu}^{\nu} \, .
\end{align}
The $c$-expansion for the $B$-field and the dilaton is given by
\bea
\hat B_{\mu \nu} &=&  - c^2 \epsilon_{a b} \tau_{\mu}{}^{a} \tau_{\nu}{}^{b} + b_{\mu \nu} \, , \\
\hat{\Phi} &=& \ln(c) + \phi  \, .
\eea
The finite Lorentz transformations after imposing $c\rightarrow \infty$ for the fundamental fields are given by
\bea
\delta_{\lambda} \tau_{\mu}{}^{a} = \lambda^{a}{}_{b} \tau_{\mu}{}^{b} \, , \qquad \delta_{\lambda} e_{\mu}{}^{a'} = \lambda^{a'}{}_{b} \tau_{\mu}{}^{b} + \lambda^{a'}{}_{b'} e_{\mu}{}^{b'} \, .
\eea
Considering the supergravity Lagrangian in string frame
\bea
S_{\rm{het}} & = & \int d^{10}x \sqrt{- \hat g} e^{-2 \hat \Phi} \left(R(\hat e) + 4 \partial_{\mu} \hat \Phi \partial^{\mu} \hat \Phi - \frac{1}{12} \hat{H}_{\mu \nu \rho} \hat{H}^{\mu \nu \rho} \right) \, ,
\label{action}
\eea
with $\hat{H}_{\mu \nu \rho}=3\partial_{[\mu} \hat B_{\nu \rho]}$
it was proved in \cite{Bergshoeff:2021bmc} that the limit $c \rightarrow \infty$ does not produce divergences.

\section{Bosonic Double Field Theory}
\label{Bosonic}
Firstly, we compute the generalised frame corresponding to the above expansion of $\hat{e}$ and $\hat{B}$. This shows that the generalised frame does not possess a finite $c$-expansion, similar to the frame $\hat{e}$ in the supergravity expansion. For this $\hat A= (A,A')$ with $A,B,... =0,\dots,3$ is a double transversal index and $A^\prime,B^\prime,...=4,\dots,2D-1$ then the generalised frame $\hat E_{M}{}^{\hat A}$ decomposes as follows:
\begin{align}
    \hat E_{M}{}^{A} &= c {E^{(1),A}_M} + \frac{1}{c} {E^{(-1),A}_M}  \, , \\
\text{with} \quad {E^{(1),A}_M} &= \frac{1}{\sqrt{2}} \left(\begin{matrix}
    - \tau_{\mu \underline{a}} - {\tau_\mu}^c \epsilon_{bc} \delta^b_{\underline a} & 0\\
    \tau_{\mu \overline{a}} - {\tau_\mu}^c \epsilon_{bc} \delta^b_{\overline a} & 0
\end{matrix} \right), \quad {E^{(-1),A}_M} = \frac{1}{\sqrt{2}} \left(\begin{matrix}
    -b_{\rho \mu} {\tau^\rho}_{\underline a} & {\tau^\mu}_{\underline a}\\
    -b_{\rho \mu} {\tau^\rho}_{\overline a} & {\tau^\mu}_{\overline a}
\end{matrix} \right)
\end{align}
and the finite
\be
\hat E_{M}{}^{A'} = \frac{1}{\sqrt{2}} 
\left(\begin{matrix}-{ e}_{\mu \underline a'}- \hat B_{ \rho\mu} {e}^{\rho }{}_{\underline a'} &  { e}^{\mu }{}_{\underline a'} \\ 
e_{\mu \overline a'}- \hat B_{\rho \mu}{} e^{\rho }{}_{\overline{a'}}& e^\mu{}_{\overline a'} \end{matrix}\right) = \frac{1}{\sqrt{2}} 
\left(\begin{matrix}-{ e}_{\mu \underline a'}- b_{ \rho\mu} {e}^{\rho }{}_{\underline a'} &  { e}^{\mu }{}_{\underline a'} \\ 
e_{\mu \overline a'}- b_{\rho \mu}{} e^{\rho }{}_{\overline{a'}}& e^\mu{}_{\overline a'} \end{matrix}\right) \, .
\ee
The gauge fixing of the double Lorentz group is given by
\bea
\tau_{\mu}{}^{\un a} \delta_{\un a}^{a} = \tau_{\mu}{}^{\ov a} \delta_{\ov a}^{a} = \tau_{\mu}{}^{a}, \qquad e_{\mu}{}^{\un a'} \delta_{\un a'}^{a'} = e_{\mu}{}^{\ov a'} \delta_{\ov a'}^{a'} = e_{\mu}{}^{\un a} \, . 
\eea
The generalised frame satisfies
\bea
\hat E_{M}{}^{\hat A} \hat \eta_{\hat A \hat B} \hat E_{N}{}^{\hat B} = \eta_{M N} 
\eea
where the invariant flat metric splits as
\be
\eta_{A B} = 
\left(\begin{matrix} - \eta_{\un a \un b} &  0 \\ 
0& \eta_{\ov a \ov b} \end{matrix}\right)  \, , \ \ \ \
\eta_{A' B'} = 
\left(\begin{matrix} - \eta_{\un a' \un b'} &  0 \\ 
0& \eta_{\ov a' \ov b'} \end{matrix}\right)  \, ,
\ee
and the O$(D,D)$ invariant metric is
\be
\eta_{M N} = 
\left(\begin{matrix} 0 &  \delta^{\mu}_{\nu} \\ 
\delta^{\nu}_{\mu} & 0 \end{matrix}\right)  \, .
\ee
The generalised metric is constructed as
\bea
\mathcal{H}_{MN} = \hat E_{M}{}^{\hat A} \hat H_{\hat A \hat B} \hat E_{N}{}^{\hat B} 
\eea
where $\hat H_{\hat A \hat B}$ is an invariant double Lorentz metric which splits as
\be
H_{A B} = 
\left(\begin{matrix}  \eta_{\un a \un b} &  0 \\ 
0& \eta_{\ov a \ov b} \end{matrix}\right)  \, , \ \ \ \
H_{A' B'} = 
\left(\begin{matrix}  \eta_{\un a' \un b'} &  0 \\ 
0& \eta_{\ov a' \ov b'} \end{matrix}\right)  \, .
\ee
The components for the generalised metric are given by
\bea
\label{genmet1}
\hat {\cal H}_{\mu \nu} & = & c^2 \tau_{\mu}{}^{a} \tau_{\nu a} + h_{\mu \nu} + \hat B_{\rho \mu} h^{\rho \sigma} \hat B_{\sigma \nu} + \frac{1}{c^2} \hat B_{\rho \mu} \tau^{\rho a} \hat B_{\sigma \nu} \tau^{\sigma}{}_{a} \, , \\
\hat {\cal H}_{\mu}{}^{\nu} & = & - \frac{1}{c^2} \hat B_{\rho \mu} \tau^{\rho a} \tau^{\nu}{}_{a} - \hat B_{\rho \mu} h^{\rho \nu}\, , \\
\hat {\cal H}^{\mu \nu} & = & \frac{1}{c^2} \tau^{\mu a} \tau^{\nu}{}_{a} + h^{\mu \nu} \, ,
\label{genmet3}
\eea
where $h_{\mu \nu}=e_{\mu}{}^{a'} e_{\nu a'}$ and $h^{\mu \nu}= e^{\mu a'} e^{\nu}{}_{a'}$. Finally, the generalised dilaton has a finite $c$-expansion, as the expansion of the dilaton $\hat{\Phi}$ and the metric determinant $\hat g$ cancel:
\bea
e^{-2 \hat{d}} = e^{-2\hat{\Phi}} \sqrt{-\hat g} = e^{-2 \phi} \sqrt{f(\tau,h)} = e^{-2 d},
\label{dilaton}
\eea 
where $f(\tau,h) = - \frac{\hat{g}}{c^4}$.

\subsection{Symmetry transformations}
The generalised frame transforms under generalised diffeomorphisms generated by $\xi^{M}$ and double Lorentz transformations generated by the parameter $\Lambda_{\hat A \hat B}$,
\bea
\delta_{\Lambda} \hat E_{M}{}^{\hat A} = {\cal L}_{\xi}\hat E_{M}{}^{\hat A} + \Lambda^{\hat A}{}_{\hat B} \hat E_{M}{}^{\hat B} \, ,
\eea
where the generalised Lie derivative for an arbitrary double vector is given by
\bea
{\cal L}_\xi V_M = \xi^{N} \partial_N V_M + (\partial_M \xi^N - \partial^N \xi_{M}) V_N + \omega \partial_{N} \xi^{N} V_{M}  \, ,
\eea 
with $\omega$ a weight factor and $\xi^{M}=(\zeta_{\mu}, \xi^{\mu})$. The previous decomposition shows that the generalised Lie derivative encodes information of the ordinary Lie derivative and, also, encodes the Abelian symmetry of the $B$-field, i.e.,
\bea
\delta_{\zeta} \hat B_{\mu \nu} = 2 \partial_{[\mu} \zeta_{\nu]} \, .
\eea
The closure of generalised diffeomorphisms requires the strong constraint, given by
\be
\partial_{M} (\partial^{M} \star) = (\partial_{M} \star) (\partial^{M} \star) = 0 \, ,
\label{SC}
\ee
where $\star$ is any DFT field/parameter. We will solve this constraint in the usual way, i.e., $\partial_{M}=(0,\partial_{\mu})$.

We split the double Lorentz parameter as
\bea
\Lambda_{A B} = \lambda_{A B} \, , \qquad \Lambda_{A A'} = \frac{1}{c} \lambda_{A A'} \, , \qquad \Lambda_{A' B'} = \lambda_{A' B'} \, .
\eea
The transformation for the components of the generalised frame is given by
\bea
\delta_{\lambda}\hat E_{M}{}^{A} & = & \lambda^{A}{}_{B} \hat E_{M}{}^{B} + \frac{1}{c} \lambda^{A}{}_{B'} \hat E_{M}{}^{B'} \, , \\
\delta_{\lambda}\hat E_{M}{}^{A'} & = & \lambda^{A'}{}_{B'} \hat E_{M}{}^{B'} + \frac{1}{c} \lambda^{A'}{}_{B} \hat E_{M}{}^{B} \, .
\eea
The closure of the double Lorentz transformations 
\bea
\Big[\delta_{\lambda_1},\delta_{\lambda_2}\Big] = \delta_{\lambda_3} \, ,
\eea
is given by the following parameters
\bea
\lambda_{3 A C} & = & 2 \lambda_{[2 A B} \lambda_{1]}^{B}{}_{C} + 2 \lambda_{[2 A B'} \lambda_{1]}^{B'}{}_{C} \, , \nn \\
\lambda_{3 A C'} & = & 2 \lambda_{[2 A B} \lambda_{1]}^{B}{}_{C'} + 2 \lambda_{[2 A B'} \lambda_{1]}^{B'}{}_{C'} \, \\
\lambda_{3 A' C'} & = & 2 \lambda_{[2 A' B'} \lambda_{1]}^{B'}{}_{C'} + 2 \lambda_{[2 A' B} \lambda_{1]}^{B}{}_{C'} \, . \nn
\eea
The gauge fixing for the components double Lorentz parameters is
\bea
\lambda_{\ov a \ov b} \delta^{\ov a}_{a} \delta^{\ov b}_{b} & = & - \lambda_{\un a \un b} \delta^{\un a}_{a} \delta^{\un b}_{b} = \lambda_{a b} \, , \nn \\
\lambda_{\ov a' \ov b'} \delta^{\ov a'}_{a'} \delta^{\ov b'}_{b'} & = & - \lambda_{\un a' \un b'} \delta^{\un a'}_{a'} \delta^{\un b'}_{b'} = \lambda_{a' b'} \, , \\
\lambda_{\ov a \ov b'} \delta^{\ov a}_{a} \delta^{\ov b'}_{b'} & = & - \lambda_{\un a \un b'} \delta^{\un a}_{a} \delta^{\un b'}_{b'} = \lambda_{a b'} \, . \nn
\eea
\subsection{Generalised fluxes and action}
The generalised fluxes are defined as
\bea
\hat F_{\hat A \hat B \hat C} & = & 3 \hat E_{[\hat A} \hat E^{M}{}_{\hat B} \hat E_{|M| \hat C]} \, , \\
\hat F_{\hat A} & = & - 2 \hat E_{\hat A} d + \sqrt{2} \partial_{M} \hat E^M{}_{\hat A} \, ,
\eea
where $\hat E_{\hat A}= \frac{1}{\sqrt 2} \hat E^M{}_{\hat A} \partial_{M}$
The relevant components of the generalised flux for constructing the action principle are: $\hat F_{\un {\hat A} \un{\hat B} \un{\hat C}}$, $\hat F_{\ov{\hat A} \un{\hat B} \un{\hat C}}$ and $\hat F_{\un{\hat A}}$. The terms in the action have the following form
\bea
\label{fluxes1}
\hat F_{\un {\hat A} \un{\hat B} \un{\hat C}} \hat F^{\un {\hat A} \un{\hat B} \un{\hat C}} & = & \hat F_{\un {A} \un{B} \un{C}} \hat F^{\un {A} \un{B} \un{C}} + 3 \hat F_{\un {A'} \un{B} \un{C}} \hat F^{\un {A'} \un{B} \un{C}} + 3 \hat F_{\un {A'} \un{B'} \un{C}} \hat F^{\un {A'} \un{B'} \un{C}} + \hat F_{\un {A'} \un{B'} \un{C'}} \hat F^{\un {A'} \un{B'} \un{C'}} \nn \\
\hat F_{\ov {\hat A} \un{\hat B} \un{\hat C}} \hat F^{\ov {\hat A} \un{\hat B} \un{\hat C}} & = & \hat F_{\ov {A} \un{B} \un{C}} \hat F^{\ov {A} \un{B} \un{C}} + 3 \hat F_{\ov {A'} \un{B} \un{C}} \hat F^{\ov {A'} \un{B} \un{C}} + 3 \hat F_{\ov {A'} \un{B'} \un{C}} \hat F^{\ov {A'} \un{B'} \un{C}} + \hat F_{\ov {A'} \un{B'} \un{C'}} \hat F^{\ov {A'} \un{B'} \un{C'}} \nn \\
\hat F_{\un {\hat A}} \hat F^{\un {\hat A}} & = & \hat F_{\un {A}} \hat F^{\un {A}} + \hat F_{\un {A'}} \hat F^{\un {A'}} \, , \nn \\
\hat E_{\un {\hat A}} \hat F^{\un {\hat A}} & = & \hat E_{\un {A}} \hat F^{\un {A}} + \hat E_{\un {A'}} \hat F^{\un {A'}} \, . \nn
\label{fluxes4}
\eea
The action principle is given by
\be
\label{DFT_action}
S = \int \ {\rm{d}}^{20}X \ e^{-2d} \ {\cal{L}}\, ,
\ee
where ${\cal L}$ is, up to total derivatives, the generalised Ricci scalar ${\cal R}$, which can be expressed in terms of the generalised fluxes as
\be
\label{generalised_ricci}
{\cal{R}}(\hat{E},d) = 2 \hat E_{\un{\hat A}}\hat F^{\un{\hat A}} + \hat F_{\un{\hat A}} \hat F^{\un{\hat A}} - \frac{1}{6}\hat F_{\un{\hat A \hat B \hat C}} \hat F^{\un{\hat A \hat B \hat C}} - \frac{1}{2}\hat F_{\ov{\hat A}\un{\hat B \hat C}} \hat F^{\ov{\hat A}\un{\hat B \hat C}}\, .
\ee
The DFT Lagrangian can be written in terms of the generalised metric as,
\bea
\hat {\cal L} & = & \frac18 \hat{\cal H}^{ M  N} \partial_{ M}\hat{\cal H}^{ K  L}\partial_{ N}\hat{\cal H}_{ K  L} - \frac12 \hat{\cal H}^{ M  N}\partial_{ N}\hat{\cal H}^{ K  L}\partial_{ L}\hat{\cal H}_{ M  K} \nn \\ && + 4 \hat{\cal H}^{ M  N} \partial_{ M}d \partial_{ N}  d - 2 \partial_{ M}\hat{\cal H}^{ M  N} \partial_{ N} d \, . \label{DFTLagrangian}
\eea
If we impose the field expansion $\hat B_{\rho \mu} =  - c^2 \epsilon_{a b} \tau_{\rho}{}^{a} \tau_{\mu}{}^{b} + b_{\rho \mu}$, the components of the generalised metric take the following form
\bea
\label{genmetB1}
\hat {\cal H}_{\mu \nu} & = & h_{\mu \nu} + b_{\rho \mu} h^{\rho \sigma} b_{\sigma \nu} + \frac{1}{c^2} b_{\rho \mu} \tau^{\rho c} b_{\sigma \nu} \tau^{\sigma}{}_{c} - 2 \epsilon_{c d}  \tau_{(\mu|}{}^{d} b_{\sigma |\nu)} \tau^{\sigma c} \, , \\
\hat {\cal H}_{\mu}{}^{\nu} & = & - \frac{1}{c^2} b_{\rho \mu} \tau^{\rho c} \tau^{\nu}{}_{c} + \epsilon_{a b} \tau_{\mu}{}^{b} \tau^{\nu a}  - b_{\rho \mu} h^{\rho \nu} \, , \\
\hat {\cal H}^{\mu \nu} & = & \frac{1}{c^2} \tau^{\mu a} \tau^{\nu}{}_{a} + h^{\mu \nu} \, .
\label{genmetB3}
\eea
Interestingly, the generalised metric can be expanded as follows,
\bea
\hat {\cal H}_{M N} =  \hat{\cal H}^{(0)}_{M N} + \frac{1}{c^2} \hat {\cal H}^{(-2)}_{M N} \, 
\eea
where
\begin{align}
    {\cal H}^{(0)}_{M N} &=  
\left(\begin{matrix} h^{\mu \nu} &  \epsilon_{a b} \tau_{\nu}{}^{b} \tau^{\mu a}  - b_{\rho \nu} h^{\rho \mu} \\ 
\epsilon_{a b} \tau_{\mu}{}^{b} \tau^{\nu a}  - b_{\rho \mu} h^{\rho \nu} &  h_{\mu \nu} + b_{\rho \mu} h^{\rho \sigma} b_{\sigma \nu} - 2 \epsilon_{c d}  \tau_{(\mu|}{}^{d} b_{\sigma |\nu)} \tau^{\sigma c} \, \end{matrix} \right) \\
    {\cal H}^{(-2)}_{M N} &=  
\left(\begin{matrix} \tau^{\mu a} \tau^{\nu}{}_{a} &  -b_{\rho \nu} \tau^{\rho c} \tau^{\mu}{}_{c} \\ 
-b_{\rho \mu} \tau^{\rho c} \tau^{\nu}{}_{c} &  b_{\rho \mu} \tau^{\rho c} b_{\sigma \nu} \tau^{\sigma}{}_{c} \end{matrix}\right)  \, .
\end{align}
At leading order, this had been demonstrated in \cite{Ko:2015rha} for flat space (Gomis-Ooguri background \cite{Gomis:2000bd}) for arbitrary backgrounds in \cite{NRDFT2}. The proporties of this Non-Riemmanian limit of DFT have been studied in \cite{NRDFT0,NRDFT1,NRDFT2,NRDFT3,Ko:2015rha}. 

The $c$-expansion components satisfy
\bea
\label{H0ODD}
{\cal H}^{(0)}_{M P} \eta^{P Q} {\cal H}^{(0)}_{Q N} & = & \eta_{M N} \\ 
{\cal H}^{(-2)}_{M P} \eta^{P Q} {\cal H}^{(-2)}_{Q N} & = & 0 \\
{\cal H}^{(0)}_{M P} \eta^{P Q} {\cal H}^{(-2)}_{Q N} & = & - {\cal H}^{(-2)}_{M P} \eta^{P Q} {\cal H}^{(0)}_{Q N} \, .
\eea
From (\ref{H0ODD}) we learn that both $\hat{\cal H}_{M P}$ and $\hat{\cal H}^{(0)}_{M P}$ are O$(D,D)$-elements. Moreover, as also the generalised dilaton is finite, $\hat{d}=d$. Hence, the DFT action is finite \cite{Ko:2015rha,NRDFT2}
\bea
\hat {\cal L}(\hat{\cal H},\hat{d}) & = & \frac18 {\cal H}^{(0) M  N} \partial_{ M}{\cal H}^{(0) K  L}\partial_{ N}{\cal H}^{(0)}_{ K  L} - \frac12 {\cal H}^{(0) M  N}\partial_{ N}{\cal H}^{(0) K  L}\partial_{ L}{\cal H}^{(0)}_{ M  K} \nn \\ && + 4 {\cal H}^{(0) M  N} \partial_{ M}d \partial_{ N}  d - 2 (\partial_{ M}{\cal H}^{ (0)M  N}) \partial_{ N} d + \mathcal{O}\left(\frac{1}{c}\right) \, . \label{cLagrangian}
\eea
The effective supergravity Lagrangian is finite, like in \cite{Bergshoeff:2021bmc}, where the authors showed this explicitly considering the expansion of the supergravity Lagrangian. From the DFT perspective, the finiteness of the Lagrangian is guaranteed. 
Taking the limit $c\rightarrow \infty$ in the double geometry is possible since $ {\cal L}^{(0)}$ is still a generalised scalar under both O$(D,D)$ and generalised diffeomorphisms, i.e., 
\bea
\delta_{\xi} {\cal L}^{(0)} = \xi^{P} \partial_{P}{\cal L}^{(0)} \, ,
\eea
and therefore the Lagrangian is well defined as NR limit of DFT. A similar analysis has been obtained in the context of Exceptional Field Theory for the membrane Newton-Cartan parameterisation \cite{Blair:2021waq}. In section \ref{Heterotic}, we will study expansions of heterotic supergravity for which such an analysis in terms of heterotic DFT is missing in the literature, so far.

\subsection{Comparison with non-Riemannian DFT}
The non-Riemannian extension for the generalised metric is given by \cite{PM,NRDFT0,NRDFT1,NRDFT2}
\begin{align}\label{paramNR}
	\mathcal{H}^{\textrm{(Non-Riem.)}}_{MN} &= \begin{pmatrix}
	 H^{\mu \nu} 	 & &  - H^{\mu \rho} {B}_{\rho \nu} + y^{\mu a} x_{\nu a} \\ - H^{\nu \rho} {B}_{\rho \mu} + y^{\nu a} x_{\mu a}
		   & &  K_{\mu\nu} - {B}_{\mu \rho} H^{\rho \sigma} {B}_{\sigma \nu} + 2 x_{(\mu| a} {B}_{|\nu)\rho} y^{\rho a}\, 
	\end{pmatrix} \, ,
\end{align}
where $K_{\mu \nu}$ and $H^{\mu \nu}$ are two symmetric tensors whose kernels are spanned by $x_{\mu}$ and $y^{\mu}$, respectively,
\bea
H^{ \mu  \nu} x_{ \nu} = K_{\mu \nu} y^{\nu} = 0 \, .
\eea
In the original formulation of \cite{PM}, the authors include also $\bar x_{\mu \bar a}$ and $\bar y^{\mu \bar a}$, which we are setting to zero from the beginning. When one compares (\ref{paramNR}) with ${\cal H}^{(0)}_{M N}$ one finds the following identifications:
\bea
K_{\mu \nu} & = & h_{\mu \nu} \, , \quad
H^{\mu \nu} =  h^{\mu \nu} \, , \nn \\
B_{\mu \nu} & = & b_{\mu \nu} \, , \\
y^{\mu a} & = & \tau^{\mu a} \, , \quad x_{\mu a} =  \epsilon_{a b} \tau_{\mu}{}^{b} \, \nn .
\eea
When comparing the full $\hat{\cal H}_{M N} =  {\cal H}^{(0)}_{M N} + \frac{1}{c^2} {\cal H}^{(-2)}_{M N}$ with the non-Riemannian generalised metric, the only difference is given by the identification
\begin{equation}
    H^{\mu \nu} =  h^{\mu \nu} + \frac{1}{c^2} \tau^{\mu a} \tau^{\nu}{}_{a}.
\end{equation}
With this, we have shown that the complete $c$-expansion of bosonic DFT in the generalised metric formalism takes very similar form the Non-Riemmanian formulation. Only $H^{\mu\nu}$ is modified and with that the property that $x$ lies in the kernel of $H$. 

\subsection{Comments about higher-derivative terms} \label{chap:HigherDer}
One important application of DFT is the construction of the four derivative terms of the bosonic and heterotic supergravity starting from an O$(D,D)$ action which schematically we can write as
\bea
S= \int d^{2D}X \ \left({\cal R}({\hat F}) + a {\cal R}^{-}(\hat F) + b {\cal R}^{+}(\hat F)\right),
\eea
and the explicit form of the action in terms of the generalized fluxes can be found in \cite{Baron:2017dvb}.  In order to find $\a'$-corrections after parametrising the fundamental fields we need to impose
 \bea
 \label{cases}
 (a,b) = 
   \begin{cases} 
      (-1,-1)  & \mbox{bosonic string \, , }  \\
      (-1,0)  & \mbox{heterotic string \, ,} \\ 
      (0,0) & \mbox{type II strings \, .} 
   \end{cases}
\eea
As initially discussed in \cite{Hohm:2016yvc}, there exists a tension between the generalised metric and frame formalism beyond the two-derivative level. Particularly, only the combination $(a,b)=(-1,1)$ can be written in terms of the generalised metric and the generalised dilaton. This theory does not correspond to a string theory formulation and it was originally constructed in \cite{Hohm:2013jaa} and further explored in \cite{Hohm:2015mka, Lescano:2016grn}. This implies that invoking the results of this section, only the HSZ theory enjoys, in principle, a finite Lagrangian using the parameterisation (\ref{genmetB1})-(\ref{genmetB3}). This means that for other choices of the parameters apart from the HSZ theory the divergences in the generalised frame (and also in the generalised fluxes) we cannot ensure, in principle, a finite structure for ${\cal R}^{-}$ and ${\cal R}^{+}$. In order to construct higher-derivative corrections from the double geometry for the NR bosonic and heterotic supergravity, we then expect modifications in the generalised Green-Schwartz transformations and, consequently, in the four derivative Lagrangian.

In the next section, we return to the two-derivative case and focus on the heterotic formulation of DFT \cite{Hohm:2011ex}.  

\section{Heterotic Double Field Theory}
\label{Heterotic}

We now consider an extended geometry with coordinates $X^{{\cal M}}=(\tilde{x}_{\mu}, x^{\mu}, x^{i})$ with ${\cal M}=0,\dots,2D-1+n$ and $i=1,\dots,n$. The generalised diffeomorphisms now contain an extra term that depends on the generalised structure constants $f_{{\cal M} {\cal N} {\cal P}}$ of the non-Abelian gauge group,
\bea
\delta_{\hat \xi} V^{ {\cal M}} = {\mathcal L}_{\hat \xi} V^{ {\cal M}} + f^{{\cal M}}{}_{ {\cal N} {\cal P}} \hat \xi^{ {\cal N}} V^{ {\cal P}}  \, ,
\label{gaugeLie}
\eea
where $V^{{\cal M}}$ is an arbitrary extended vector. The structure constants are fully skewsymmetric and satisfy the Jacobi identity,
\bea
f_{{ {\cal M} {\cal N} {\cal P}}}=f_{[{\cal M} {\cal N}  {\cal P]}}\, , \qquad f_{[ {\cal M} {\cal N}}{}^{{\cal R}}f_{{ {\cal P}}] {{\cal R}}}{}^{{\cal Q}}=0\, . \label{consf}
\eea
The closure of the algebra of generalised diffeomorphisms requires the strong constraint (\ref{SC}) plus an extra constraint given by
\bea
f_{{\cal M} {\cal N} {\cal P}} \partial^{\cal M} \star = 0 \, .
\label{Strongf}
\eea
This new constraint can be solved when the generalised structure constants are non-vanishing only for ${\cal M}, {\cal N}, ...=i,j,k,...$, i.e. $f_{{\cal M} {\cal N} {\cal P}} \equiv f_{ijk}$, and $\partial_{i}=0$. The parameterisation of the invariant metric is 
\be
{\eta}_{{{\cal M} {\cal N}}}  %= \left(\begin{matrix}\eta^{\mu\nu}&\eta^\mu{}_\nu&\eta^\mu{}_i\\ 
%\eta_\mu{}^\nu&\eta_{\mu\nu}&\eta_{\mu i}\\\eta_{i}{}^\nu&\eta_{i\nu}&\eta_{ij}\end{matrix}\right)
= \left(\begin{matrix}0&\delta^\mu{}_\nu&0\\ 
\delta_\mu{}^\nu&0&0\\0&0&\kappa_{ij}\end{matrix}\right) \ , \label{eta}
\ee
with $\mu, \nu,... =0,..., D-1$, $i,j,...=1,..., N$ and $\kappa_{ij}$ the Killing metric of the gauge group which can be used to freely raise and lower the $i,j,\dots$-indices.

\paragraph{Expansions of the vector field.} Whereas the expansion of the metric seems to be universal, this does not appear to be true for vector gauge fields. Even Maxwell theory in the ordinary Galilean NR limit has three distinct limits: \cite{GalileanEM1,GalileanEM2}
\begin{itemize}
    \item \textit{electric limit}, $A \rightarrow c \, a^{(\tau)} + a^{(h)}$
    \item \textit{magnetic limit}, $A \rightarrow a^{(\tau)} + c a^{(h)}$
    \item \textit{special} or \textit{Galilean electrodynamics limit} $A \rightarrow c \, a_1^{(\tau)} + \frac{1}{c} a_2^{(\tau)} + a^{(h)}$
\end{itemize}
$a^{(\tau)}$, $a^{(h)}$ denote components along the projections defined by $\tau$ and $h$ respectively. 

Heterotic supergravity is even more involved: it contains a non-Abelian gauge field $A$ and its particular interplay with the metric and $B$-field via the Chern-Simons 3-form. At this point, many limits are possible. Let us present three (stringy) NR expansions that account for our central incentive: a \textit{finite supergravity action} in the limit $c \rightarrow \infty$.
\begin{itemize}
    \item \textit{Bergshoeff-Romano expansion} \cite{BergshoeffRomano}: $A \rightarrow c^2 a^{(\tau)} + a^{(h)}$

    This expansion is a non-trivial expansion of all involved supergravity fields, also necessitating a modification of the metric expansion $g \sim c^4 g^{(\tau)} + c^2 g^{(\tau)} + g^{(h)}$. In section \ref{chap:BRexpansion} we demonstrate that this expansion does not lead to a finite generalised metric, although it leads to a finite supergravity action \cite{BergshoeffRomano}.

    \item \textit{'Natural' 1-form expansion}: $A \rightarrow c \ a^{(\tau)} + \frac{1}{c} a$

    This is a new suggestion for the expansion of $A$, similar to the one from Galilean electromagnetism. The leading order in $c$ can also be understood as a natural expansion of a 1-form. To see that, consider an expansion of $A$ in flat coordinates 
\begin{equation}
	A = A_{\hat{a}} \hat{e}^{\hat{a}} = A_a \hat{e}^a + ... \rightarrow c A_a \tau^{a} + ...
\end{equation}    
Following this logic, a fundamental $(0,p)$-tensor scales with $c^p$. This is also consistent with the $c^2$-scaling of metric and $B$-field. 

It will be shown in section \ref{chap:finiteExpansion} that this expansion together with usual expansion for metric and $B$-field leads to a \textit{finite generalised metric} and heterotic DFT (and supergravity) action, given a particular choice for $a^{(\tau)}$.

    \item \textit{Trivial expansion}: $A \rightarrow a$

This case, together with the expansion of the NS-NS fields from section \ref{Review} give a finite generalised metric and a finite heterotic DFT. In this case the gauge field cannot be treated as a 1-form at the supergravity level. At leading order, this would be the non-Riemannian parameterisation of heterotic DFT given in \cite{NRDFT3}. 

\end{itemize}

\subsection{The Bergshoeff-Romano expansion} \label{chap:BRexpansion}
In \cite{BergshoeffRomano}, the authors start by expanding the vielbein as
\bea
\hat e_{\mu}{}^{-} = c \ \tau_{\mu}{}^{-}, \quad \hat e_{\mu}{}^{+} =  c \ \tau_{\mu}{}^{+} - \frac{c^3}{2} \alpha^2_{-} \tau_{\mu}{}^{-}\, \label{modification}, \quad \hat e_{\mu}{}^{a'} = e_{\mu}{}^{a'} ,
\eea
where $v^{\pm}=\frac{1}{\sqrt{2}}(v^{0} \pm v^{1})$. The inverse vielbein is given by
\bea
\label{modification2}
\hat e^{\mu}{}_- & = \frac{1}{c} \tau^{\mu}{}_-, \quad \hat e^{\mu}{}_+ = \frac{1}{c} \tau^{\mu}{}_+ + \frac{c}{2} \alpha^2_{-} \tau^{\mu}{}_-, \quad \hat e^{\mu}{}_{a'} = e^{\mu}{}_{a'} .
\eea
The gauge field $\hat A_{\mu}{}^i$ is expanded as
\bea
\hat{A}_\mu^i = c^2 \tau_{\mu}{}^{-} \alpha^{i}_- + \tau_{\mu}{}^{+} \alpha^{i}_+ + e_{\mu}{}^{a'} a_{a'}{}^i .
\eea
Also, % in (\ref{modification}) and (\ref{modification2}),
 it is understood that $\alpha^2_{-} = \alpha_{-i} \alpha_-^i$ and $\alpha_{+-} = \alpha_{+i} \alpha_-^i$. The $c$-expansions for the $B$-field and the dilaton are given by
\bea
\hat{B}_{\mu \nu} &=& - c^2 \epsilon_{a b} \tau_{\mu}{}^{a} \tau_{\nu}{}^{b} (1 + \alpha_{+-}) + 2 c^2 \tau^{-}{}_{[\mu} e_{\nu]}{}^{a'} \alpha_{-a'} + b_{\mu \nu}  \, , \label{eq:BerghoeffRomanoBField} \\
\hat{\Phi} &=& \ln(c) + \phi \, .
\eea
The field content of this expansion consists of the NS-NS stringy Newton-Cartan fields $\tau,h,b,\phi$ and the vector-valued 1-form $(\alpha,a)$, one component of which is singled out to give the leading $c^2$-contribution. According to \cite{BergshoeffRomano}, this expansion is motivated by finiteness of the limit $c \rightarrow \infty$ of the heterotic supergravity action, of local Lorentz and $B$-field gauge transformations and of the longitudinal $T$-duality transformations. Let us briefly discuss the first point, for discussion of the symmetries we refer to \cite{BergshoeffRomano}.

Considering the heterotic supergravity Lagrangian in string frame, 
\bea
S_{\rm{het}} & = & \int d^{10}x \sqrt{- \hat g} e^{-2 \hat \Phi} \left( R(\hat e) + 4 \partial_{\mu} \hat \phi \partial^{\mu} \hat \phi - \frac{1}{12} \hat{\bar{H}}_{\mu \nu \rho} \hat{\bar{H}}^{\mu \nu \rho} - \frac{1}{4} \hat{F}_{\mu \nu} \hat{F}^{\mu \nu} \right) \, ,
\label{actionhet}
\eea
with 
\be
\hat{\bar{H}}_{\mu\nu\rho}=3\left(\partial_{[\mu} \hat{B}_{\nu\rho]}-\hat{C}_{\mu\nu\rho}^{(f)}\right)\, , \label{barH}
\ee 
and $C_{\mu\nu\rho}^{(g)}$ is the Chern-Simons 3-form, defined as
\be
\hat{C}_{\mu\nu\rho}^{(f)}= \hat{A}^i_{[\mu}\partial_\nu \hat{A}_{\rho]i}-\frac13 \hat{f}_{ijk} \hat{A}_\mu^i \hat{A}_\nu^j \hat{A}_\rho^k \, , 
\ee
it was proved in \cite{BergshoeffRomano} that the limit $c \rightarrow \infty$ does not produce divergences. As in the bosonic case this happens due to cancellations between divergences of different terms in \eqref{actionhet}. 

\paragraph{Embedding in heterotic DFT.} We start by promoting the NR of the vielbein in terms of the generalized frame of heterotic DFT. In order to do so, we need to split the double Lorentz index as $\hat{\cal A}= (m,p,{\cal A}')=(m,p,A',\bar i)$, where $m=(\un m, \ov m)$, $p=(\un p,\ov p)$, $A=4,\dots,2D-1$ and $\bar i=1,\dots,N$. The conventions for the flat metric in the $-,+$ directions are $\eta_{-+}=\eta^{-+}=-1$, $\eta_{+-}=\eta^{+-}=-1$, while the conventions for the epsilon tensor are $\epsilon_{-+}=-\epsilon_{+-}=1$.   

The generalised frame can be decomposed  as
\begin{align}
    E^{\hat {\cal M}}{}_{m} \ &= \
\frac{1}{\sqrt{2}}\left(\begin{matrix}-c{ \tau}_{\mu \underline m} - \frac{1}{c} \hat{C}_{ \rho\mu} {\tau}^{\rho }{}_{\underline m} &  \frac{1}{c}{\tau}^{\mu }{}_{\underline m} & \frac{1}{c} {\alpha}_{\underline{p}}{}^i \, , \\ 
c{\tau}_{\mu \overline m} - \frac{1}{c} \hat{C}_{ \rho\mu} {\tau}^{\rho }{}_{\overline m} & \frac{1}{c}{\tau}^{\mu }{}_{\ov m} & \frac{1}{c} {\alpha}_{\overline{p}}{}^i   \end{matrix}\right)  \nn \, , \\
E^{\hat {\cal M}}{}_{p}\ &= \
\frac{1}{\sqrt{2}}\left(\begin{matrix}-c{ \tau}_{\mu \underline p} + \frac{c^3}{2} {\alpha}^2_{-} \tau_{\mu \un m} -  \hat{C}_{ \rho\mu} (\frac{1}{c}{\tau}^{\rho }{}_{\underline p} + \frac{c}{2} {\alpha}^2_{-} \tau^{\rho}{}_{\un m}) &  \frac{1}{c}{\tau}^{\mu }{}_{\underline p} + \frac{c}{2} {\alpha}^2_{-} \tau^{\mu}{}_{\un m}  & c {\alpha}_{\underline{m}}{}^i - c {\alpha}^2_{-} {\alpha}_{\underline p}{}^{i}\, , \\ 
c{\tau}_{\mu \overline p} - \frac{c^3}{2} {\alpha}^2_{-} \tau_{\mu \ov m} -  \hat{C}_{ \rho\mu} (\frac{1}{c} {\tau}^{\rho }{}_{\overline p} + \frac{c}{2} {\alpha}^2_{-} \tau^{\rho}{}_{\ov m}) & \frac{1}{c}{\tau}^{\mu }{}_{\ov p} + \frac{c}{2} {\alpha}^2_{-} \tau^{\mu}{}_{\ov m} & c {\alpha}_{\overline{m}}{}^i - c {\alpha}^2_{-} {\alpha}_{\overline p}{}^{i}  \end{matrix}\right)  \, , \nn \\
E^{\hat {\cal M}}{}_{\cal A'} \ &= \
\frac{1}{\sqrt{2}}\left(\begin{matrix}-{ e}_{\mu \underline a'}-\hat{C}_{ \rho\mu} { e}^{\rho }{}_{\underline a'} &  { e}^{\mu }{}_{\underline a'} & -\hat{A}_{\nu}{}^i e^{\nu \un a'} \, , \\ 
e_{\mu \overline a'}-\hat{C}_{\rho \mu}{} e^{\rho }{}_{\overline{a'}}& e^\mu{}_{\overline a'}&-\hat{A}_{\nu}{}^i e^{\nu \ov a'}  \\
\sqrt2 \hat{A}_{\mu i}e^i{}_{\overline i} - \frac{1}{\sqrt 2} c^2 {a}_{+i}{\alpha}^2_{-} (\tau_{\mu}{}^{+} + \tau_{\mu}{}^{-})e^i{}_{\overline i}  &0&\sqrt2 e^i{}_{\overline i} \end{matrix}\right) \label{genfr}  \,,
\end{align}
where $\hat{C}_{\mu\nu}=\hat{B}_{\mu\nu}+\frac12 \hat{A}_{\mu}^i \hat A_{\nu i}$, $e^i{}_{\overline i}$  the inverse vielbein for the Killing metric of the $SO(32)$ or $E_8 \times E_8$  gauge group, $\eta_{ij}=e_i{}^{\overline i} \eta_{\overline i\overline j}e_j{}^{\overline j}$,  as required for  modular invariance of the heterotic string. The gauge fixing condition is given by $\tau_{\mu \un m} \delta^{\un m}_{m} = \tau_{\mu \ov m} \delta^{\ov m}_{m} = \tau_{\mu -}$, $\tau^{\mu}{}_{\un m} \delta^{\un m}_{m} = \tau^{\mu}{}_{\ov m} \delta^{\ov m}_{m} = \tau^{\mu}{}_{-}$, $\tau_{\mu \un p} \delta^{\un p}_{p} = \tau_{\mu \ov p} \delta^{\ov p}_{p} = \tau_{\mu +}$, $\tau^{\mu}{}_{\un p} \delta^{\un p}_{p} = \tau^{\mu}{}_{\ov p} \delta^{\ov p}_{p} = \tau_{\mu +}$,  $e_{\mu \ov a'} \delta^{\ov a'}_{a'} = e_{\mu \un a'} \delta^{\un a'}_{a'} = e_{\mu a'}$ and $e^{\mu \ov a'} \delta_{\ov a'}^{a'} = e^{\mu \un a'} \delta_{\un a'}^{a'} = e^{\mu a'}$, ${\alpha}_{\un m i}={\alpha}_{\ov m i}= {\alpha}_{-i}$, ${\alpha}_{\un p i}={\alpha}_{\ov p i}= {\alpha}_{+i}$. 

The generalised frame (\ref{genfr}) is compatible with the O$(D,D+n)$ generalised metric, and using \eqref{modification} and \eqref{modification2}, the latter
\begin{align}
    \hat{\cal H}_{\cal M N} = \left(\begin{matrix} \hat{g}^{\mu \nu} & - \hat{g}^{\mu \rho} \hat{C}_{\rho \nu} & - \hat{g}^{\mu \rho} \hat{A}_{\rho}^j \\
- \hat{g}^{\nu \rho} \hat{C}_{\rho \mu} & \hat{g}_{\mu \nu} + \hat{C}_{\rho \mu} \hat{C}_{\sigma \nu} \hat{g}^{\rho \sigma} + \hat{A}_{\mu}{}^k \kappa_{kl} \hat{A}_{\nu}{}^l &
\hat{C}_{\rho \mu} \hat{g}^{\rho \sigma} \hat{A}_{\sigma}^j + \hat{A}_{\mu}{}^j \\
- \hat{g}^{\nu \rho} \hat{A}_{\rho}^i & \hat{C}_{\rho \nu} \hat{g}^{\rho \sigma} \hat{A}_{\sigma}^i +  \hat{A}_{\nu}{}^i & \kappa^{ij} + \hat{A}_{\rho}^i \hat{g}^{\rho \sigma} \hat{A}_{\sigma}^j \end{matrix}\right),
\label{eq:HeteroticGeneralisedMetric}
\end{align}
 is not finite in the limit $c \rightarrow \infty$. It is straightforward to compute the explicit form of the generalised metric, which has an expansion with $c^4$- and $c^2$-divergences:
\bea
&{}&  \hat{\cal H}_{\cal M N}(\hat{g},\hat{B},\hat{A}) \label{eq:BRExpansionGeneralisedMetric}\\
&=& c^4 \ {\cal H}^{(4)}_{\cal M N} (\tau,h,b,\alpha,a)  + c^2 \ {\cal H}^{(2)}_{\cal M N}(\tau,h,b,\alpha,a)  + {\cal H}^{(0)}_{\cal M N}(\tau,h,b,\alpha,a) + \mathcal{O}\left(\frac{1}{c^2}\right) \, . \nonumber
\eea
A component to explicitly show this problem is
\bea
{\cal H}^{i j} & = & \kappa^{i j} + {a}^{a'}{}^{i} {a}_{a'}^j - 2 {\alpha}_{-}^{(i} {\alpha}_{+}^{j)} - 2 c^2 {\alpha}^2_{-} {\alpha}_{+}^{(i} {\alpha}_{-}^{j)} \, .
\eea
%One could be tempted to keep this form of the generalised metric and evaluate the Lagrangian, since the important condition is that the latter must be finite. 
From here, one can evaluate the heterotic DFT action, whose Lagrangian has the form \cite{Hohm:2011ex}
\begin{equation}
    \mathcal{L} = \mathcal{R}(\hat{\mathcal{H}},d) - \mathcal{R}_{\hat{f}}(\hat{\mathcal{H}},\hat{f})
\end{equation}
with
\begin{align}
   \mathcal{R}(\hat{\mathcal{H}},d) &= 4 \hat{\mathcal{H}}^{ \cal MN} \partial_{\cal M} \partial_{\cal N} d - \partial_{\cal M} \partial_{\cal N} \mathcal{H}^{\cal MN} - 4 \hat{\mathcal{H}}^{\cal MN} \partial_{\cal M} d \partial_{\cal N} d + 4 \partial_{\cal M} \hat{\mathcal{H}}_{\cal MN} \partial_{\cal N} d \label{eq:HetDFTAct1}\\
   &{} \quad + \frac{1}{8} \hat{\mathcal{H}}^{\cal MN} \partial_{\cal M} \hat{\mathcal{H}}_{\cal KL} \partial_{\cal N} \hat{\mathcal{H}}^{\cal KL} - \frac{1}{2} \hat{\mathcal{H}}^{\cal M N} \partial_{\cal M} \hat{\mathcal{H}}^{\cal KL} \partial_{\cal K} \hat{\mathcal{H}}_{\cal LN} \nonumber \\
   \mathcal{R}_{\hat{f}}(\hat{\mathcal{H}},{\hat{f}}) &= \frac{1}{2} {{\hat{f}}^{\cal K}}{}_{\cal MN} \hat{\mathcal{H}}^{\cal MP} \hat{\mathcal{H}}^{\cal NQ} \partial_{\cal P} \hat{\mathcal{H}}_{\cal QM} \label{eq:HetDFTAct2}  \\
   &{} \quad + \frac{1}{12} {\hat{f}}^{\cal M}{}_{\cal KP} {\hat{f}}^{\cal N}{}_{\cal LQ} \hat{\mathcal{H}}_{\cal MN} \hat{\mathcal{H}}^{\cal KL} \hat{\mathcal{H}}^{\cal PQ} + \frac{1}{4} {{\hat{f}}^{\cal M}}{}_{\cal NK} {{\hat{f}}^{\cal N}}{}_{\cal ML} \hat{\mathcal{H}}^{\cal KL} + \frac{1}{6} {\hat{f}}^{\cal MNK} {\hat{f}}{}_{\cal MNK} \ . \nonumber
\end{align}
%At the DFT level, the Lagrangian under the proposal of Bergshoeff and Romano is divergent. 
When imposing the strong constraint, this object reduces to the heterotic supergravity action \eqref{actionhet}, which is finite in this parameterisation as explained in \cite{BergshoeffRomano}. At the DFT level, however, this parameterisation contains $c^4$- and $c^2$-divergences, which lead to $c^4$-divergences for $\mathcal{R}(\hat{g})$ and $\hat{\bar H}_{\mu \nu \rho}$, and $c^2$-contributions to $\mathcal{R}(\hat{g})$, $\hat{\bar H}_{\mu \nu \rho}$ and $\hat{F}_{\mu \nu i}$. This means that we cannot uplift a NR heterotic supergravity to the DFT formalism, because we need $c^2$- and $c^4$-contributions to construct the T-duality multiplets. By construction, the full $\hat{\mathcal{H}}$ is an element of O$(D,D+n)$. But, in contrast to the bosonic case, this is not the case for the leading order $\mathcal{H}^{(4)}$ by itself. 

Since this expansion works at the supergravity level but is not fully compatible with the O$(D,D)$ symmetry, we propose an alternative expansion for the heterotic degrees of freedom in the following section. 

\subsection{An alternative $c$-expansion} \label{chap:finiteExpansion}

In contrast to the previous section, here we will derive a $c$-expansion for the heterotic supergravity fields from the assumptions of \textit{finiteness of the generalised metric} for $c \rightarrow \infty$. Arranging the degrees of freedom into heterotic DFT (or generalised geometry) will ensure the correct transformation behaviour of generalised diffeomorphisms including $B$-field gauge transformations and non-Abelian gauge transformations. For this, the gauge symmetry needs to obtain an expansion as well. In turn, the fact the $T$-duality of such a NR heterotic background leads to another NR heterotic background is automatic.

Consider the following ansatz:
\begin{align}
    \hat{g}_{\mu \nu} &= c^2 {\tau_\mu}^a {\tau_\nu}^b \gamma_{ab} + h_{\mu \nu}, \nonumber\\
    \hat{B}_{\mu \nu} &= c^2 {\tau_\mu}^a {\tau_\nu}^b \beta_{ab} + b_{\mu \nu}, \label{eq:HeteroticExpansion1}\\
    \hat{A}_{\mu} &= c \ {\tau_\mu}^a \alpha_a^i + \frac{1}{c} a_\mu^i . \nonumber
\end{align}
This allows for some freedom to account for differences to the bosonic case, in metric and $B$-field expansion via $\beta_{ab}$ and $\gamma_{ab}$, which are skewsymmetric and symmetric $2 \times 2$-matrices respectively. The inverse metric is still given by $\hat{g}^{\mu\nu} = h^{\mu \nu} + \frac{1}{c^2} {\tau_a}^\mu {\tau_b}^\nu \gamma^{ab}$. 

This NR heterotic limit consists of the following field content: Besides the usual stringy NR fields $\tau,h,b$ and $\phi$ there is the gauge field with a divergent part $\alpha$ along the $\tau$-directions and, similar to the $B$-field expansion, an unconstrained part $a$ at order $\frac{1}{c}$. This parameterisation makes it seem that the components $a_\mu^i$ are not relevant at leading order and, hence, this parameterisation might oversimplify things. This is, in fact, not true as the in heterotic supergravity the gauge field also contributes through the Chern-Simons terms, in the generalised metric are captured by $\hat{A}_\mu^i \hat{A}_{\nu i}$-terms:
\begin{align}
    \frac{1}{2} \hat{A}_\mu^i \hat{A}_{\nu i} = c^2 {\tau_\mu}^a {\tau_\nu}^b \alpha_{ab} + a_{\mu \nu} + \frac{1}{c^2} a^{(2)}_{\mu \nu}
\end{align}
with $\alpha_{ab} = \frac{1}{2} \alpha_a^i \alpha_{b i}$, $a_{\mu \nu} = {\tau_{(\mu}}^a  a_{\nu)}^i \alpha_{ai}$ and $a^{(2)}_{\mu \nu} = \frac{1}{2} a_\mu^i a_{\nu i}$. This also means that one cannot choose $\alpha =0$ without decoupling the gauge field at leading order. For $\hat{C}_{\mu \nu} = \hat{B}_{\mu\nu} + \frac{1}{2} \hat{A}_\mu^i \hat{A}_{\nu i}$ this implies
\begin{equation}
   \hat{C}_{\mu \nu} = c^2 {\tau_\mu}^a {\tau_\nu}^b ( \alpha_{ab} + \beta_{ab} ) + (a_{\mu \nu}+  b_{\mu \nu} ) + \frac{1}{c^2} a^{(2)}_{\mu \nu}.
\end{equation}
Remarkably, the heterotic generalised metric \eqref{eq:HeteroticGeneralisedMetric} is finite, given appropriate choices of $\alpha_a^i$, $\beta_{ab}$ and $\gamma_{ab}$. The crucial components with potential divergences are $\mathcal{H}_{\mu \nu}$ and ${\mathcal{H}_{\mu}}^i$:
\begin{align}
\label{system1}
    \hat{\cal{H}}_{\mu\nu} &= c^2 {\tau_\mu}^a {\tau_\nu}^b \left( \gamma_{ab} +2 \alpha_{ab} + (\alpha_{ac} - \beta_{ac}) \gamma^{cd} (\alpha_{db} + \beta_{bd}) \right) + \mathcal{O}(c^0) \overset{!}{=} 0 + \mathcal{O}(c^0) \\
    {\hat{\cal{H}}_\mu}{}^i &= c \ {\tau_\mu}^a \left( ( \alpha_{ac} - \beta_{ac} ) \gamma^{cb} + \delta_a^b \right) \alpha_b^i + \mathcal{O}(c^{-1}) \overset{!}{=} 0 + \mathcal{O}(c^{-1}) \, .
    \label{system2}
\end{align}
This is a system of $6$ cubic equations with the unknowns $\alpha_a^i$, $\beta_{ab}$ and $\gamma_{ab}$. This system seems difficult to solve in generality, but one can find two simple solutions with one free choice of gauge vector $\alpha_\pm^i$ each:
\begin{align}
    \alpha^i &= (\alpha_+^i , \alpha_-^i) \equiv (\alpha_+^i , 0), \quad \beta_{ab} = \epsilon_{ab}, \quad \gamma_{ab} = \eta_{ab} \nonumber \\
    \text{or} \qquad \alpha^i &= (\alpha_+^i , \alpha_-^i) \equiv  (0 , \alpha_-^i), \quad \beta_{ab} = -\epsilon_{ab}, \quad \gamma_{ab} = \eta_{ab} . \label{eq:NewExpansionSolution}
\end{align}
\paragraph{The generalised metric.} With this expansion,
\begin{align}
    \hat{g}_{\mu \nu} &= c^2 {\tau_\mu}^a {\tau_\nu}^b \eta_{ab} + h_{\mu \nu} , \quad \hat{B}_{\mu \nu} &= \pm c^2 {\tau_\mu}^a {\tau_\nu}^b \epsilon_{ab} + b_{\mu \nu} \label{eq:HeteroticExpansion}, \quad \hat{A}_{\mu}^{(\pm)} &= c\ {\tau_\mu}^\pm \alpha_\pm^i + \frac{1}{c} a_\mu^i, 
\end{align}
the components of the generalised metric look as follows:
\begin{align}
   &{} \qquad \hat{\cal H}_{\cal M N}(\hat{g},\hat{B},\hat{A}) \label{eq:HeteroticGeneralisedMetricFinite}  \\
   &= {\mathcal{H}}^{(0)}_{\cal MN} (\tau,h,b,\alpha,a) + \frac{1}{c} {\mathcal{H}}^{(-1)}_{\cal MN} (\tau,h,b,\alpha,a) + \frac{1}{c^2} {\mathcal{H}}^{(-2)}_{\cal MN} (\tau,h,b,\alpha,a) + \mathcal{O}\left( \frac{1}{c^3} \right) \nonumber
\end{align}
where\footnote{The expressions would simplify if ${\tau_a}^\mu a_\mu^i = 0$, i.e. if these degrees of freedom are transversal. For example, the $\mathcal{O}\left(\frac{1}{c^3}\right)$-terms of the generalised metric would vanish. These longitudinal parts could considered to be $\frac{1}{c^2}$-corrections to $\alpha_a^i$, in analogy to the inclusion of the Bargmann-like field ${m_\mu}^a$ as $\frac{1}{c^2}$-correction of ${\tau_\mu}^a$. Latter are important when considering the NR limit of the string worldsheet \cite{Bergshoeff:2019pij}.}
\begin{align*}
    \hat{\cal H}^{\mu\nu} &= h^{\mu\nu} + \frac{1}{c^2} {\tau_a}^\mu {\tau_b}^\nu \eta^{ab} \\
   {\hat{\cal H}^\mu}{}_\nu &= - \left(h^{\nu \rho} {c}_{\rho \mu} + {\tau_\mu}^a {\tau_b}^\nu \eta^{bc} (\alpha_{ca} + \beta_{ca}) \right) - \frac{1}{c^2} \left( c_{\rho \mu} {\tau_a}^\rho {\tau_b}^\nu \eta^{ab} + h^{\nu \rho} a_{\rho \mu}^{(2)} \right) + \mathcal{O}\left(\frac{1}{c^4}\right) \\
    {\hat{\cal H}^{\mu i}} &= - \frac{1}{c} \left( {\tau_a}^\mu \alpha_b^i \eta^{ab} + h^{\mu \rho} a_\rho^i \right) + \mathcal{O}\left(\frac{1}{c^3}\right) \\
    {\hat{\cal H}_{\mu\nu}} &= \left( h_{\mu\nu} + 2 a_{\mu\nu} + {c}_{ \rho \mu} h^{\rho \sigma} {c}_{\sigma \nu} + 2 c_{\rho(\mu} {\tau_{\nu)}}^c {\tau_a}^\rho \eta^{ab} (\alpha_{bc} + \beta_{bc}) \right) \\
    &{} \quad + \frac{1}{c^2} \left( c_{\kappa \mu} c_{\lambda \nu} {\tau_a}^\kappa {\tau_b}^\lambda \eta^{ab} + 2 a^{(2)}_{\rho(\mu} {\tau_{\nu)}}^c {\tau_a}^\rho \eta^{ab} (\alpha_{bc} + \beta_{bc})  \right) + \mathcal{O}\left(\frac{1}{c^4}\right) \\
    {\hat{\cal H}_{\mu}}{}^i &= \frac{1}{c} \left( c_{\rho \mu} {\tau^\rho}_a \alpha_b^i \eta^{ab} + \left( c_{\rho \mu} h^{\rho \sigma} + {\tau_\mu}^a {\tau_c}^\nu (\alpha_{ab} - \beta_{ab}) \eta^{bc} + \delta^\nu_\mu \right) a_\nu^i \right) + \mathcal{O}\left(\frac{1}{c^3} \right) \\
    \hat{\cal H}^{ij} &= \left( \kappa^{ij} + \eta^{ab} \alpha_a^i \alpha_b^j \right) + \frac{1}{c^2} \left( h^{\mu \nu} a_\mu^i a_\nu^i + 2 {\tau_a}^\mu a_\mu^{(i} \alpha_b^{j)} \right) + \mathcal{O}\left(\frac{1}{c^4} \right)
\end{align*}
with $c_{\mu\nu} = b_{\mu \nu} + a_{\mu \nu}$. For the 'null solutions' for $\alpha$ \eqref{eq:NewExpansionSolution}, $\eta^{ab} \alpha_a^i \alpha_b^j = 0$. This simplifies the expression for $\hat{\mathcal{H}}^{ij}$.

At leading order, the generalised metric is
\begin{align}    
    {\mathcal{H}}^{(0)}_{\cal MN} &= \begin{pmatrix}
	 h^{\mu \nu} 	 &  - h^{\mu \rho} {c}_{\rho \nu} + y^{\mu}_a x_{\nu}^a & 0 \\ - h^{\nu \rho} {c}_{\rho \mu} + y^{\nu}_a x_{\mu}^a
		   &   h_{\mu\nu} + 2 a_{\mu\nu} - {c}_{\rho \mu} h^{\rho \sigma} {c}_{\sigma \nu} + 2  {c}_{\rho (\mu} x_{\mu)}^a y^{\rho}_a & 0 \\ 0 & 0 & \kappa^{ij} 
	\end{pmatrix}, \label{eq:HeteroticGMLeadingOrder}
\end{align}
with $x_\mu^a = {\tau_\mu}^a$ and $y^\mu_c = - {\tau_a}^\mu \eta^{ab} (\alpha_{bc} + \beta_{bc})$. This could be labeled as a \textit{non-Riemannian parameterisation of the heterotic string}. It fits into a modification of the non-Riemannian ansatz \eqref{paramNR} by non-skewsymmetric $b_{\mu \nu} \rightarrow c_{\mu \nu}$. It trivially extends to $n$ gauge directions. Also, it includes additional $a^2$-terms in $\mathcal{H}_{\mu\nu}$. $x_\mu^a$ and $y^\mu_c$ both lie in the kernel of $h^{\mu \nu}$ resp. $h_{\mu\nu}$. It differs from \cite{NRDFT3} in that the linear terms in the gauge field appears at orders $\frac{1}{c}$.

\paragraph{Two expansions for the structure constants.} Using the generalised Lie derivative for the heterotic DFT \eqref{gaugeLie} 
it is possible to read off the gauge transformations for the NR degrees of freedom. In order to ensure that the generalised metric \eqref{eq:HeteroticGeneralisedMetricFinite} transforms correctly under generalised diffeomorphisms \eqref{gaugeLie}, the gauge parameter needs to be expanded as
\begin{equation}
    \hat{\xi}_{\mathcal{M}} = \left(\xi_\mu , \xi^\mu, \frac{1}{c} \lambda^i \right).
\end{equation}
There are different choices for expansions of the structure constants $\hat{f}_{ijk}$, leading to different gauge transformations and expansions of the field strength $F$ and the Chern-Simons 3-form $C^{f}$:
\begin{itemize}
    \item $\underline{\hat{f}_{ijk} = c f_{ijk}}$:

    In this case, the field strengths are:
    \begin{align}
        \hat{F}_{\mu\nu}^k &= c \ 2 \nabla^{(a)}_{[\mu} \left({\tau_{\nu]}}^a \alpha_a^k \right) + \frac{1}{c} F^{(a),k}_{\mu\nu} + \mathcal{O}\left( \frac{1}{c^3} \right) \\
        \hat{C}^{(\hat{f})} &= 2 \ c^2 \tau^a \wedge \mathrm{d} \tau^b \alpha_{ab} - f_{ijk} \alpha_a^i \ \tau^a \wedge a^j \wedge a^k + \mathcal{O}\left( \frac{1}{c^2} \right)
    \end{align}
    where $\nabla^{(a)}_\mu = \partial_\mu + [ \cdot , a_\mu ]$ is the covariant derivative and $F^{(a)}$ the usual non-Abelian field strength for the sub-leading connection $a$. The gauge transformations take a natural form,
    \bea
    \delta_{\lambda} a_{\mu}^k & = & \partial_{\mu} \lambda^{k} + f^{k}{}_{ij} \lambda^{i} a_{\mu}^j = \nabla^{(a)}_\mu \lambda^k ,\label{connection} \\
    \delta_{\lambda} \alpha_a^{i} & = & f^{i}{}_{jk} \lambda^{j} \alpha_a^{k} \, , \label{vector} \\  
    \delta_{\lambda} b_{\mu \nu} & = & - \partial_{[\mu} \lambda^{i} \tau_{\nu]}^a \alpha_{a i} \, , 
    \label{GS}
    \eea
    while the remaining fields $\tau,h,\phi$ are gauge invariant. The transformation (\ref{connection}) suggests that $a_{\mu i}$ acts as the physical gauge connection, (\ref{vector}) shows that $\alpha_{\pm}^{i}$ transforms as a vector and (\ref{GS}) is a NR Green-Schwarz mechanism. 
    
    In order to obtain the above form of the field strengths and gauge transformations, the concrete form of \eqref{eq:NewExpansionSolution} has to be used, in particular $f_{ijk} \alpha_a^j \alpha_b^k = 0$.
    
   Despite the straightforward interpretation of the NR fields $\alpha$ and $a$, this choice has two caveats:
    \begin{itemize}
        \item The finiteness of the heterotic DFT action \eqref{eq:HetDFTAct1} \& \eqref{eq:HetDFTAct2} is \textit{not automatic}. Whereas \eqref{eq:HetDFTAct1} is finite due to the finiteness of generalised metric and generalised dilaton, $\mathcal{R}_{\hat{f}}$ in \eqref{eq:HetDFTAct2} might contain divergences due to the $\hat{f} \hat{\mathcal{H}}^2$- and $\hat{f}^2 \hat{\mathcal{H}}$-terms. Nevertheless, due to the particular form of the generalised metric \eqref{eq:HeteroticGeneralisedMetricFinite} and the structure constants $f_{ijk}$, one notices that
        \begin{align*}
            \mathcal{R}_{\hat{f}}(\hat{\mathcal{H}},\hat{f}) = \mathcal{O}\left( \frac{1}{c^2} \right).
        \end{align*}
        Hence, the DFT action is finite in this expansion.

        \item The gauge algebra becomes singular for $c\rightarrow \infty$:
        \begin{align}
            [\hat{\xi}^i , \hat{\xi}^j] = {\hat{f}}_k {}^{ij} \hat{\xi}^k \quad \Rightarrow \quad [\lambda^i , \lambda^j] = c^2 {{f}}_k {}^{ij} \lambda^k . 
        \end{align}
        \end{itemize}
        
    \item $\underline{\hat{f}_{ijk} = \frac{1}{c} f_{ijk}}$:

    In this case the interpretation of the expansions of the field strength and gauge transformation is more obscure. For example, for latter one gets:
        \bea
    \delta_{\lambda} a_{\mu}^k & = & \partial_{\mu} \lambda^{k} + 2 f^{k}{}_{ij} \lambda^{i} \tau_{\mu}^a \alpha_{a}^k \label{connection2} \\
    \delta_{\lambda} \alpha_{\pm}^{i} & = & 0 \, , \label{vector2} \\  
    \delta_{\lambda} b_{\mu \nu} & = & - \partial_{[\mu} \lambda^{i} \tau_{\nu]}^a \alpha_{a i} \, , 
    \label{GS2}
    \eea
    So, neither does $a$ transform as non-Abelian gauge connection nor $\alpha$ as vector under gauge transformation. On the other hand, in this expansion we have that
    \begin{itemize}
        \item the finiteness of the heterotic DFT action \eqref{eq:HetDFTAct1} \& \eqref{eq:HetDFTAct2} is \textit{manifestly} guaranteed, due to the finiteness of generalised metric and generalised dilaton.

        \item the gauge algebra is non-singular for $c\rightarrow \infty$:
        \begin{align}
            [\hat{\xi}^i , \hat{\xi}^j] = {\hat{f}}_k {}^{ij} \hat{\xi}^k \quad \Rightarrow \quad [\lambda^i , \lambda^j] = {{f}}_k {}^{ij} \lambda^k . 
        \end{align}
        \end{itemize}
\end{itemize}

\section{Discussion}
\label{Discussion}
In this work we presented a NR expansion for the degrees of freedom of DFT and generalised geometry, for both the bosonic and the heterotic case. In the former, the expansion of the generalised metric is convergent and the NR supergravity Lagrangian can be recovered only from the leading order ${\cal H}^{(0)}_{M N}$, meaning that the $c\rightarrow \infty$ limit can be taken in double geometry. This setup is in agreement with the non-Riemannian formulation of DFT even if the NR limit is taken at the DFT level. In this paper we consider the generalised metric formulation of DFT as a fundamental theory, and the supergravity framework as a solution to the strong constraint. However, from the opposite perspective, when the supergravity theory is the starting point and we want to rewrite it in the double geometry, it is possible to capture the dynamics of the finite NR supergravity using T-duality multiplets. This means that it is possible to construct a generalised metric Lagrangian considering only the supergravity contributions of the Lagrangian after taking the NR limit. In the heterotic case, however, this is not possible when $\hat{A}_{\mu}^i \sim c^2$. This means that it is possible to construct the NR supergravity from the heterotic DFT setup as we showed in section \ref{Heterotic} but the NR limit has to be taken after solving the strong constraint and breaking the duality group. We cannot reverse engineer in this case and rewrite the finite NR heterotic Lagrangian in the double geometry because of the need of divergent terms which are required in the double geometry. Nevertheless, in this work we propose an alternative expansion in order to avoid this problem. Our proposal consist in a $\hat{A}_{\mu}^i \sim c$ expansion, which leads to a finite generalised metric in the framework of heterotic DFT. Therefore, we ensure that the supergravity action is finite from the double geometry construction. 

The present article continues a series of works \cite{NRDFT0,NRDFT1,NRDFT2,NRDFT3,Blair:2021waq,Berman:2019izh} that highlight the usefulness of DFT and generalised geometry in understanding NR limits of supergravity. In particular, arranging degrees of freedom of supergravity in NR limits in terms of the generalised metric of DFT makes is tractable how cancellations of divergences happens. The generalised metric also appears naturally in the Hamiltonian formulation of the string worldsheet \cite{Duff:1989tf,Tseytlin:1990nb,Tseytlin:1990va,Siegel1,Siegel2,Alekseev:2004np,Osten:2019ayq}. This was extended to the heterotic world-sheet in \cite{Hatsuda:2022zpi,Osten:2023cza,Hassler:2023nht} and to general worldvolumes in string and $M$-theory in the context of Exceptional Field Theory in \cite{Hatsuda:2012vm,Hatsuda:2013dya,Osten:2021fil,Hatsuda:2023dwx,Osten:2024mjt}. Hence, one could expect that $c$-expansions in DFT or Exceptional Field Theory are useful when taking the (string or $p$-brane) NR-limits of these worldvolume theories. As for the point particle in Newton-Cartan geometry or the string in String Newton-Cartan geometry, it is expected that a more general expansion, including $\frac{1}{c^2}$-corrections of the $\tau$, is necessary for this \cite{Bergshoeff:2019pij}.

In the context of heterotic supergravity, the results of this work provide a framework to construct alternative expansions for the gauge field, the $B$-field and the vielbein, directly from the double geometry solving the system (\ref{system1})-(\ref{system2}), since we cannot ensure that our solution is unique. A natural continuation is to deeply explore the NR limit of heterotic supergravity and construct the NR heterotic supergravity action explicitly. The inclusion of matter such as the fermionic field content coming from the double geometry \cite{Susydft1,Susydft2,Susydft3}, or more general matter \cite{EN1,EN2,EN3} to connect with stringy formulations of hydrodynamics/thermodynamics preserving the duality invariance and cosmological set-ups \cite{Cosmo1,Cosmo2,Cosmo3,Cosmo4,Cosmo5}, are other contexts where the NR limit could be addressed using similar techniques as the one used here.

\vspace*{-15pt}
\subsection*{Acknowledgements}
\vspace*{-5pt}
We are grateful to Chris D.A. Blair for comments on the first version and helpful remarks on the development of the finiteness of the non-relativistic limit of bosonic DFT in the literature. E.L. thanks Mariana Graña for enlighting discussions in the first stages of this work during his visit Université Paris-Saclay covered by the ERC Consolidator Grant 772408-Stringlandscape. E.L. is supported by the SONATA BIS grant 2021/42/E/ST2/00304 from the National Science Centre (NCN), Poland. The work of D.O. is part of project No. 2022/45/P/ST2/03995 co-funded by the National Science Centre (NCN) and the European Union’s Horizon 2020 research and innovation programme under the Marie Sk\l odowska-Curie grant agreement no. 945339.

\vspace*{10pt}
\includegraphics[width = 0.09 \textwidth]{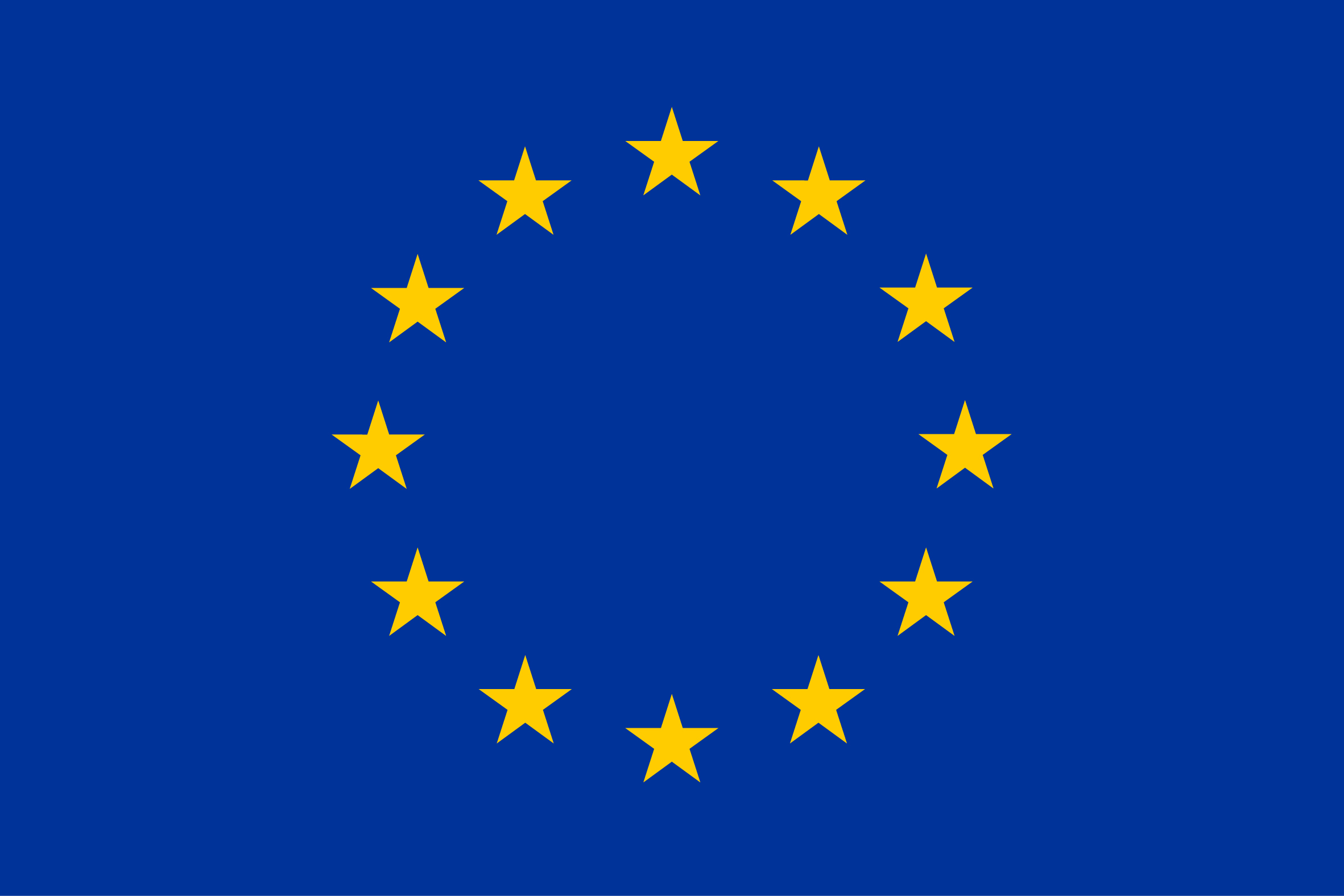} $\quad$
\includegraphics[width = 0.7 \textwidth]{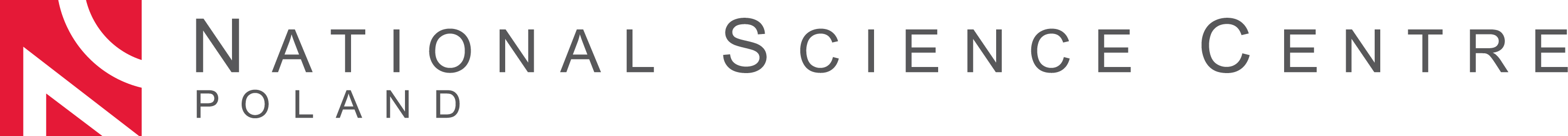}

\bibliographystyle{jhep}
\bibliography{ReferencesNR}
\end{document}